\documentclass[12pt, prd,superscriptaddress,nofootinbib, preprintnumbers]{revtex4}

\usepackage{amsmath}
\usepackage{amsfonts}
\usepackage{graphicx}

\begin{document}

\title{\Large A NEW FUNDAMENTAL DUALITY IN NUCLEI\\
AND ITS IMPLICATIONS FOR\\
QUANTUM MECHANICS}

\author{S. Afsar Abbas\footnote{afsar.ctp@jmi.ac.in}}
\affiliation{Centre for Theoretical Physics\\
JMI, New Delhi - 110025, India }

\vspace{1in}

\begin{abstract}
The Liquid Drop Models (LDM) and the Independent Particle Models
(IPM) have been known to provide two conflicting pictures of the
nucleus. The IPM being quantum mechanical, is believed to provide a
fundamental picture of the nucleus and hence has been focus of the
still elusive unified theory of the nucleus. It is believed that the
LDM at best is an effective and limited model of the nucleus. Here,
through a comprehensive study of one nucleon separation energy, we
give convincing evidence that actually the LDM is as fundamental and
as basic for the description of the nucleus as the IPM is. As such
the LDM and the IPM provide simultaneously co-exiting complementary
duality of the nuclear phenomena. This fundamental duality also
provides solution to the decades old Coester Band problem in the
nucleus. Similarity and differences with respect to the well known
wave-particle duality, as envisaged in Bohr's Complementarity
Principle, is pointed out. Thereafter implications of this new
Duality in the nucleus for quantum mechanics is discussed.

\end{abstract}

\maketitle

\newpage

In the search for a "unified theory" of nuclear physics, in spite of
intense efforts in the last 70 years or so,  the dream remains unfulfilled.
A large number of models have been proposed to study the nucleus in specific
situations and in which cases these are quite successful. A single
comprehensive theory of the nuclear phenomena however is still lacking.

Broadly speaking, all these large number of models can be classified
as being of two kind: Liquid Drop Models (LDM) and Independent Particle
Models (IPM). There is a basic and underlying conflict between these two:
as the LDM requires a strongly interacting nucleus displaying a
classical liquid drop character, while the IPM consists of weakly
interacting and independent particles in a shell and which is quantum
mechanical in character.
Whether these diverse models are of the LDM kind or the IPM kind is
determined by the underlying initial philosophy which goes into building
the Hamiltonian of the model.
Thereafter incorporated are various kind of corrections:
primarily of the IPM compensating kind if
the initial model was LDM, and primarily of the LDM
compensation kind if the model was IPM initially.
This is the broad picture of all the large number of models
in nuclear physics [1]. Henceforth we shall use the acronym LDM and IPM
to classify all these large number of nuclear models in use today.
As IPM is basically quantum mechanical in character, it is felt that
the putative "unified theory" would be intrinsically of IPM in
character with suitable correction terms. This dream still remains
unfulfilled in spite of intense efforts in this direction.
The LDM is believed to provide at best a tentative and approximate picture
of the nucleus. In fact many a "purist" would have it banished but for
the fact that it offers a simple and straightforward description of
several effects in nuclei. However, can we take the fact of the failure
to obtain a viable "unified theory" of the nucleus as of now, as hinting at
the possibility that perhaps the LDM intrinsically has "more" in it than
what it has been credited for?

To understand the basis and the relationship between the LDM and IPM, we
study the single neutron separation energies $S_{N}$ and the single proton
separation energies $S_{p}$ in depth. In fact, as already well known,
successful fitting of  $S_{N}$ as a function of neutron numbers N
( for various fixed proton numbers Z )
and of fitting $S_{p}$ as a function of Z ( for various fixed N ), has
been one of the initial checks on the validity of a particular model.
As it turns out practically all the models come out successful in this
regard. We have checked this within various models ( indicated below ) and
confirm that by and large all these models are very successful in plotting
separation energies as indicated above - that is
$S_{N}$ vs N (for various fixed Z, both magic and non-magic)
and $S_{p}$ vs Z (for various fixed N, both magic and non-magic).
The success of all these models as to the above separation energies
indicates that if we were to plot separation energies slightly
differently - that is:
$S_{N}$ vs Z (for various fixed N) and $S_{p}$ vs N (for various fixed Z),
then too we should get equally successful fits for all these models.
We have checked that this too is always true ( hundreds of cases - both
magic and non-magic )
{\bf EXCEPT} for just three cases (to be discussed below).
There is no a priori reason for the failure in just these three cases
while the same models are so successful in fitting {\bf ALL} the
the separation energies plotted for any N or any Z and plotted in any
cosistent manner.
This unexpected failure of separation energies in just these three cases
is very puzzling.

The experimental data on binding energies and various separation energies
is available in Ref. [2]. As a representation of the LDM we take the most
comprehensive and the most respected compilation of their model
calculations of the so called FRDM in Ref. [3].
Within the IPM the most respected and widely quoted is the comprehensive
compilation of the calculation of mean field theory model of HFBCS-1 [4]
(which we call HFB in short).
These two are the most successful and widely accepted versions
of the LDM and the IPM categories respectively.
As one more representation of the IPM
category we also use the published compilation of the so called INM [5].

We now have comprehensive plotting of
$S_{N}$ vs N ( for various fixed Z - both magic and non-magic );
$S_{N}$ vs Z ( for various fixed N - both magic and non-magic );
$S_{p}$ vs Z ( for various fixed N - both magic and non-magic );
and $S_{p}$ vs N ( for various fixed Z -  both magic and non-magic ).
Plotted were the experimental numbers and compared with
FRDM, HFB and INM model calculation results.
As expected we find that all these models are by and large pretty much
successful in fitting all these separation energies.
This gives credence to the statement that all these are
"successful" models of the nucleus.
We find that ONLY for three cases there is a marked failure of the
HFB and INM cases while FRDM is still successful in these cases also.
These are: $S_{N}$ vs Z for N=126 and N=82
and $S_{p}$ vs N  for Z=82. We plot these in Figs. 1,2 and 3 respectively.
The success of FRDM even in these three cases attests to the
validity of basic underlying philosophy of liquid drop character of
the nucleus with suitable IPM kind of corrections.
Is the failure of the IPM models pointing towards the
shortcoming of having ignored the liquid drop character of the nucleus?

For $S_{N}$ vs Z for N=$N_{magic}$, the core consists of
$N_{magic} + Z_{magic}$ . Now as more protons are added, the core would be
affected say due to the mean field created by the extra protons each
time. This is the well known effect in shell model; e.g. the
electromagnetic properties of $O^{17}$ can only be explained by
incorporating effective charges induced due to core polarization as a
result of
the extra neutron sitting outside the $O^{16}$ core. Now in the above
separation energy we are picking up one neutron from this polarized core.
This core remains the same as Z goes from $Z_{min}$ to $Z_{max}$
( both of which are given by the empirical data ).

Now any self consistent successful calculation in IPM ( e.g. HFBCS-1 )
should have reproduced the $S_{N}$ vs Z for N=$N_{magic}$
at N=82 and 126 respectively. It works for all the different neutron
numbers and so it should have self consistently reproduced the above
separation energy. So any such polarization should have been reproduced
self-consistently by HFBCS-1. The same parameters of HFBCS-1
which successfully fit all the other separation energies ( of the above
kind as well ), just cannot be changed arbitrarily for these two nuclei
only. Hence clearly this failure should be taken as a fundamental
conundrum. This suggests that, unkwnowingly or knowingly,
we are missing some fundamental aspect of the nuclear reality,
and which is manifesting itself only here in terms of the separation
energies. What is it that this IPM kind of HFBCS-1 is missing?
Quite clearly all possible conceivable corrections have, one way or the
other, been included in these IPM kind of models.

We know that IPM definitely ignores the liquid
drop character of the nucleus. It was never taken as reflecting any
fundamental reality of the nucleus.
But it may just be that this assumption
was wrong and that the solution to this
conundrum may lie in a proper incorporation of the LDM character
in the IPM.

To understand the issue under discussion,
we plot a schematic shell structure in Fig. 4.
We assume that the neutron
number is fixed at $N_{magic}$ so to say at N=82 and 126 respectively
for Figs 1 and 2. For proton numbers, first one sees it as closed at
$Z_{magic}$ so as to form a doubly magic closed core.
This discussion would correspond to the plots in Figs. 1 and 2
of $S_{N}$ vs Z for N=82 and N=126 respectively.
That is one adds more protons on top of the magic number as the Z number
is changed. So for the N=82 and 126 cases, $Z_{magic}$ is 50 and 82
respectively ( see figs. 1 and 2 ).
We can treat the doubly magic core of all these nuclei for different
Z as forming the liquid drop kind of core. For different Z numbers this
core would be successively deformed or polarized more and more.
In these separation energies we are pulling out the same single neutron
for different Z numbers. Now as stated above we expect all these
polarization effects to be taken care of automatically in any IPM.
As expected IPM works well for all, except these three cases.
Clearly it is not because of any fundamental shortcoming of
these IPM, and hence, we are forced to conclude, that this may be
arising from the ignored liquid drop character in these models.

Let us study Fig 1 and 2 carefully. First
the HFB plot with respect to the experimental plot in Fig. 1.
We notice that there is a clear pattern in the manner
in which HFB is missing the experimental points. First for
Z=81 the HFB point is opposite in direction to the experimental point.
Next right from Z=82 upwards the HFB plot runs parallel to the
experimental plot almost at a constant distance from it. This constant
difference between the two lines is significant and shows that it is
independent of the Z number. On the basis of figure 4, it suggests that
this constant difference must be a function of the liquid drop core, a
fixed number for all Z numbers, that is of

$A_{magic}=N_{magic}+Z_{magic}$

which is 126+82=208 for this case. We know that in liquid drop models,
in the binding energy the most significant term is
a volume term ( contributing the dominant $a_{v} A$ term
to the binding energy with $a_{v}$ = 16 MeV ).
So here too we notice that the almost constant
difference of about 0.5 MeV between the theory and the experiment
can be understood to be arising from a volume term
$b_{v} A_{magic}$ which is due to
the liquid drop like core for all these
nuclei. For the case under study with $A_{magic}$ = 208
we get $b_{v}$ = 0.0024 MeV.
Note that this LDM kind of correction is, clearly beyond the periphery
of all the corrections incorporated in HFBCS-1 ( or any other IPM for
that matter ).

Next the situation for N=82 case ( Fig.2 ) is the same with about a
similar value for $b_{v}$. For the other IPM model considered here,
that is the INM, it also similarly misses the experimental points
completely and systematically ( with no
structure whatsoever ), in the same manner as the HFB model does
( in fact more clearly and cleanly! ).
This too supports the discussion above indicating here also
the necessity of and LDM kind of correction.

Now let us study $S_{p}$ vs N for Z=82 (Fig. 3).
Here too FRDM fits the data
quite well. This supports, as discussed earlier, the basic underlying
LDM philosophy of FRDM. To understand the manner in which the IPM
are failing here, we need to have a figure like Fig. 4 but now
exchanging symbols N and Z. Note that HFB has even opposite behaviour at
N=106, 121 and 125. Most interestingly it fits the data nicely
for N=126 and above. So given magic Z=82 fixed here, the
relevant neutron magic number is N=126.
Hence the "core" is doubly magic A=82+126=206.
Thus the lower N numbers in the figure may be considered as "holes"
in this liquid drop like "core".
Hence, as discussed above the almost constant difference between the HFB
data and experiment below N=126 is due to the LDM effect
with about the same $b_{v}$ value.
The INM is also missing the data in about the same manner as discussed
above. Therefore the same conclusion as to the role of LDM is relevant
here too. Hence we conclude that to explain the three anomalies,
it is necessary to include both IPM and LDM simultaneously.

If this be so, then this co-existence of the LDM and IPM
characteristic should obviously
be present in all the nuclei, light and heavy.
However we find that for smaller magic numbers like N/Z=20, 50 etc
there are no prominent effects like in the above three heavy nuclei
cases. We plot these in Figs. 5 and 6.
At best one may note some pre-cursor effects, like for example
for the INM cases in Figs. 5 and 6.

This indicates that though, the above IPM+LDM effect must be
present in all nuclei, it does not manifest itself that clearly and
prominenetly for lighter nuclei. For lighter nuclei apparently
the sensitivity of the liquid drop rigidity is not so prominent
so as to stand out above what may be parametrized in any good IPM.
What we notice is that this liquid drop becomes rigid enough
to stand out above whatever the best IPM models can parametrize,
only for heavy nuclei.
The sensitivity and freedom of good parametrization ( as apparently
done in HFB etc) is good enough to mask and liquid drop effects that
should be present in light nuclei as well. However, for heavy nuclei,
even the best parametrization fails to mask the same,
wherein there appears as a much stronger liquid drop effect.
The separation energy, plotted in the special manner, as done here,
has allowed us a peep through a narrow window at this unique
phenomenon.

Note that we found that though present in any nuclei, light or heavy, it
was actually getting manifested clearly in separation energy studies,
only for heavier nuclei.
Hence one expects that for Infinite Nuclear Matter the
presence of LDM character should be even more manifest in any IPM study.
Hence it should have something to say about a resolution
of the Coester Band problem.

Indeed, this Duality also provides an answer to the well known several
decades old puzzle, the Coester Band problem [6].
It is known that the best realistic
interactions ( like the Reid Interaction ), using the best techniques,
cannot reproduce the Nuclear Matter properties. They miss the binding
energy and the saturation density of Nuclear Matter in a systematic
manner, along the so called Coester Band. It is found [6] that when the
saturation density is fitted, the binding energy comes out too low
( as compared to the 16 MeV value ). Clearly these results fall within the
ambit of our definition of the IPM. Hence on the basis of our
discussion here, all these IPM are ignoring the LDM effects at a
fundamental level.
The corrections in the Coester Band are therefore clearly arising
due to the non-inclusion of the LDM part.

The above discussion makes it clear that the resolution of the separation
energy anomalies in the three (clear cut) cases is the unambiguous
inclusion of the LDM character in the IPM approximation.This liquid drop
property is beyond any IPM approaches with a correction of any other kind
or an approximation of any other kind.It is an additional effect which
shows itself when all that conceivably should and could have been done,
has been successfully executed. Nuclear physics is teaching us that the
IPM approach has an unanticipated limitation.
One can only get a complete description of the nuclear phenomenon within
IPM by including the orthogonal liquid drop characteristics as well.
IPM is quantum mechanical while the LDM is classical in character.
The fact that these two - classical and quantum, are appearing side by
side with equal fundamental relevance is a surprise. But it is clear that
the two co-exist simultaneously and in a complementary manner to provide a
complete description of the nucleus.

Hence this is pointing to a fundamental "Duality" in nuclear physics.
The two apparently conflicting points of view - the LDM (classical in
character) and IPM (quantum mechanical in character)
are providing dual description of the nucleus. They actually co-exist and
are required simultaneously in a complementary manner to provide a
complete and consistent description of the nucleus. Hence this is a
fundamental duality in physics.

We have found here that
the LDM is not an approximation of an IPM of any kind.
And also that neither is it
some kind of an effective model of the nucleus.
It is as fundamental in describing nucleus as an IPM is.
In fact, both of these are simulataneosly needed to describe the nucleus.
As such these are complementary in nature to provide a
new Duality in nuclear physics.

Hence we suggest that the nuclear phenomena should schematically
be represented as


$\langle {\it nucleus} \rangle = \langle {\it IPM} \rangle +
  \langle {\it LDM} \rangle$


This new Duality between the classical Liquid Drop Model and the quantum
Independent Particle Model of the nucleus is very reminiscent
of the wave-particle duality of photon and matter particles
in quantum mechanics. Though reminiscent, it is quite independent
of the same.

The wave-particle duality has been a source of much confusion;
while de Broglie thought that actually it is " wave {\it AND}
particle " while Bohr argued for " wave {\it OR} particle " [7].
Today as part of the Copenhagen interpretation, Bohr's view is
more dominant. Bohr's  point of view is dubbed as the Complementarity
Principle [7]. The wave and particle aspects of a quantum entity
is exclusive in nature, as per Bohr.

This exclusiveness is an essential aspect of the Complementarity Principle.
Bohr believed that there was a single quantum reality of say a photon or
an electron, and due to classical measurement, it reduces to two distinct
classical realities - that of a wave nature or a particle nature. Now as
logically it would be inconsistent for a single quantum reality
to manifest itself simultanaeously as two conflicting classical
properties of wave and particle nature. Hence if this were the whole
truth then there is nothing wrong with Bohr's logic - that the wave and
the particle aspects be exclusive of each other.
Experimentally there has been support of Bohr's Complementarity
Principle of exclusive wave- particle duality and this is well documented
[7]. However, ever since the de Broglie suggestion and later due to Bohm's
work, efforts have been around to find photon and matter particles like
electron manifesting both wave and particle characters simultaneously
in a single experiment [7]. One should mention the suggestion of a
suitable such experiment by Ghose, Home and Agarwal [8]
and its claimed experimental confirmation by Mizobushi and Ohtoke [9].
They show that actually both the wave and particle aspects
of photon and matter particles may co-exist simultaneously.
One should also mention the recent claims as to same effect by Afshar [10].

Now as we have found a new Duality in nuclei here in this paper,
wherein  both the LDM and IPM co-exist.
This is the only duality known outside the wave-particle duality
of quantum mechanics. There would be a conflict between these two
dualities, as per the Complementarity Principle of Bohr the duality is
exclusive , while in the nuclear Duality, the two co-exist simultaneously.
However as discussed above, there are convincing experimental evidences
that Bohr's exclusive Complementarity Principle may be wrong.
In that case, there is no conflict between the
wave-particle duality and the nuclear IPM-LDM duality.

Should quantum mechanics manifest itself at macroscopic distances?
First obvious answer would be "no"! However there are several well known
phenomena like superconductivity and superfluidity where clearly quantum
phenomena is distinctly exhibited at macroscopic scales. So as to explain
the Meissner effect it is necessary to assume that the macroscopic wave
function is unperturbed by the magnetic field. So to say, it becomes
"rigid"! In recent years the so called "macroscopic quantum mechanics"
has become popular due to the ongoing studies on Josephson
junction. Macroscopic tunneling through the Josephson junction is
routinely observed [11].
Even well above the classical-quantum crossover temperature,
the macroscopic quantum effects in Josephson junction has been observed
[12]. McDonald [13] has studied the sharpness of the quantized energy
levels in two Josephson junctions and he states,
" The fact that the myriad of interactions of ${10}^{12}$ electrons in a
macroscopic body, a Josephson junction, can produce sharply defined energy
levels, suggests a dynamical state effectively divorced from the
complexity of its environment. The existence of this state, the
macroscopic quantum state of superconductor, is well established."

This just consolidates that actually as shown in this paper, the basic
physical reality may be


$\langle {\it reality} \rangle = \langle {\it micro} \rangle
  + \langle {\it macro/classical} \rangle$


It may happen that in a particular situation only micro
quantum aspect dominates ( as in say atomic physics )
and in another where only the macro quantum reality dominates
( as say in superconductivity ). However, in some cases it may happen that
the macroscopic part above may appear as
entirely to be, as what we may be willing to call as "classical".
For example, as we have observed in nuclear physics here, where the
micro part is the IPM and the macro/classical part is the LDM.
It seems that this may happen when both the micro and micro
are co-existing in a physical phenomena - just as we have found
here in the nucleus.
This just blurs the distinction of what is macro ( meaning classical
really ) and what is micro.

Deep reflections on the measurement dilemma has been forcing physicists
to think beyond the deep divide between
the micro and the macro as envisaged in
Bohr's Complementarity Principle. The speculation that there may be some
unknown but intrinsic connection between the macro and the micro has
actually been around for some time. For example as per Bell [14, p. 171],
" Now in my opinion the founding fathers were in fact wrong on this
point. The quantum phenomena do not exclude a uniform description of micro
and macro world .. systems and apparatus. It is not essential to
introduce a vague division of the world of this kind. This was indicated
already by de Broglie in 1926, when he answered the question:
"wave or particle?" by "wave and particle" ."

What has only been speculated above, as shown in this paper,
has actually found concrete
expression in terms of a new Duality between the LDM and the IPM
degrees of freedom in nuclei.
Hence there is no division between the micro and the macro/classical.
In fact they are part of the unity of Nature. Both of these
co-existing simultaneously in a fundamental manner, is what the nucleus
is teaching us.

ACKNOWLEDGEMENT: I am grateful to Ratna Abbas for discusions and
Farooq Bhat for plotting the figures.

\newpage

\begin{figure}[ht]
\centering
\includegraphics[totalheight=0.7\textheight, angle=-90]{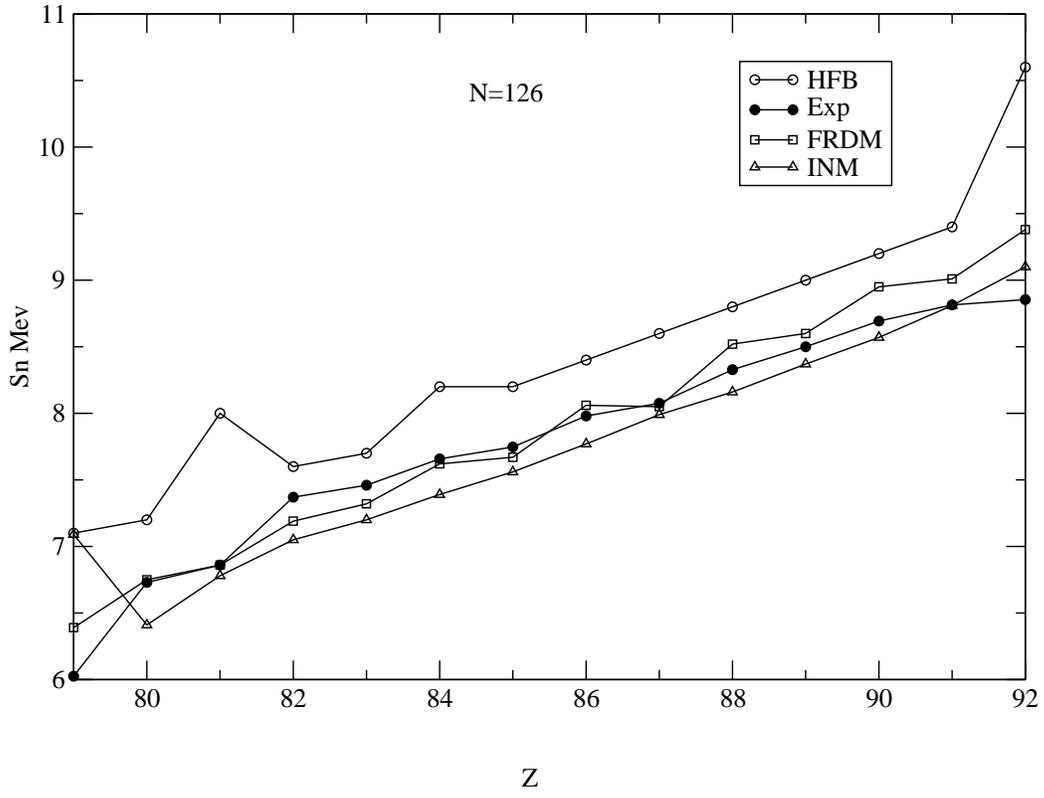}
\caption{$ S_{N} $ vs Z for fixed N=126 }
\end{figure}

\newpage

\begin{figure}[ht]
\centering
\includegraphics[totalheight=0.7\textheight, angle=-90]{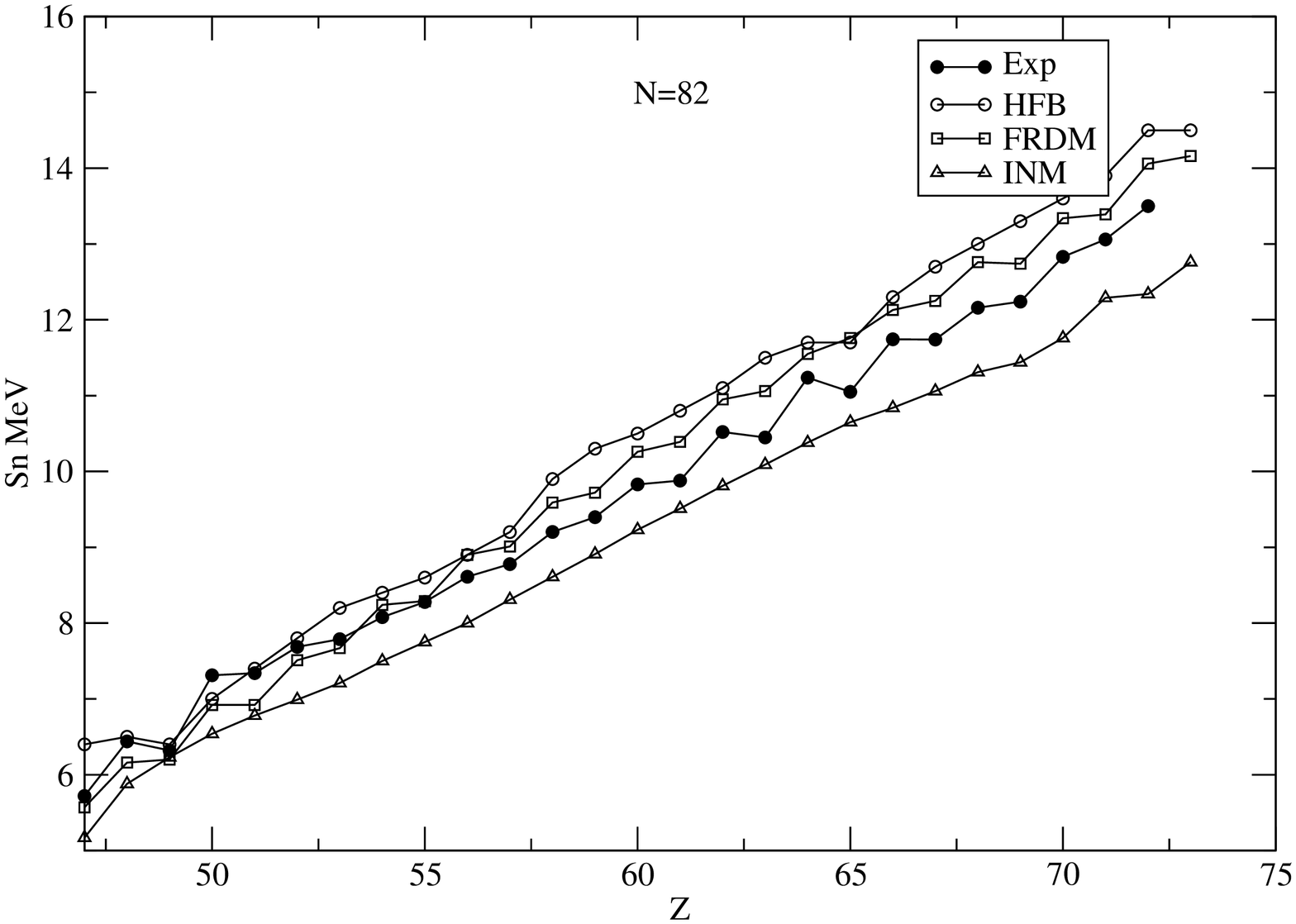}
\caption{$ S_{N} $ vs Z for fixed N=82}
\end{figure}

\newpage

\begin{figure}[ht]
\centering
\includegraphics[totalheight=0.7\textheight, angle=-90]{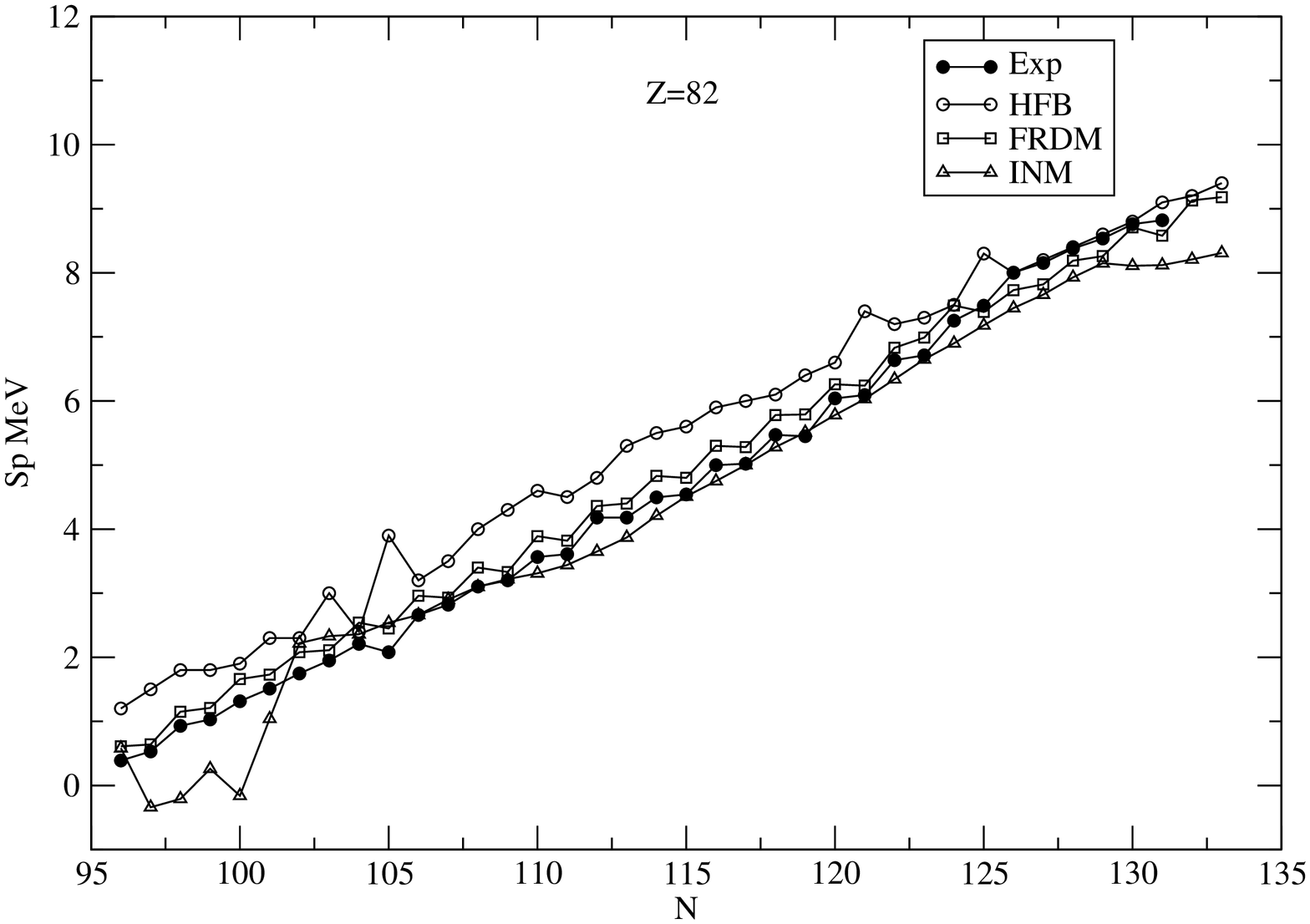}
\caption{$ S_{p} $ vs N for fixed Z=82}
\end{figure}

\newpage

\begin{figure}[ht]
\centering
\includegraphics[totalheight=0.5\textheight, angle=0]{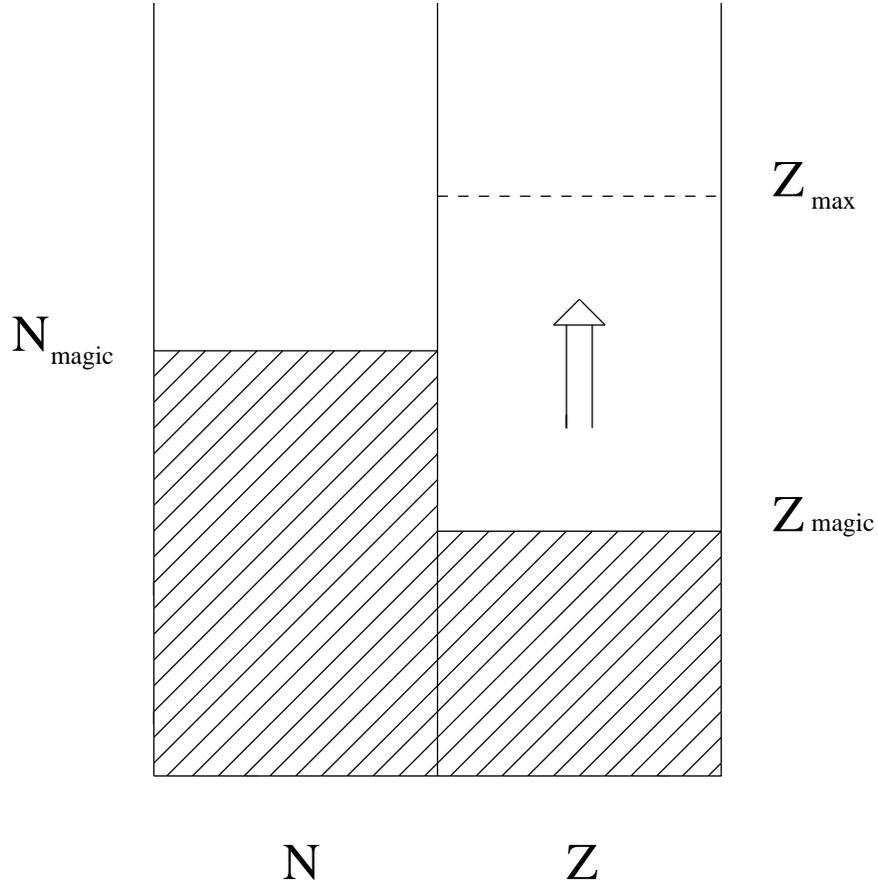}
\caption{ Schematic shell model for N and Z. This is for
understanding $ S_{N} $ vs Z for N=82 and 126. As Z is changed, the
core for any Z is $Z_{magic}$ (Z including $Z_{max}$ is given by the
empirically existing data)}
\end{figure}

\newpage

\begin{figure}[ht]
\centering
\includegraphics[totalheight=0.7\textheight, angle=-90]{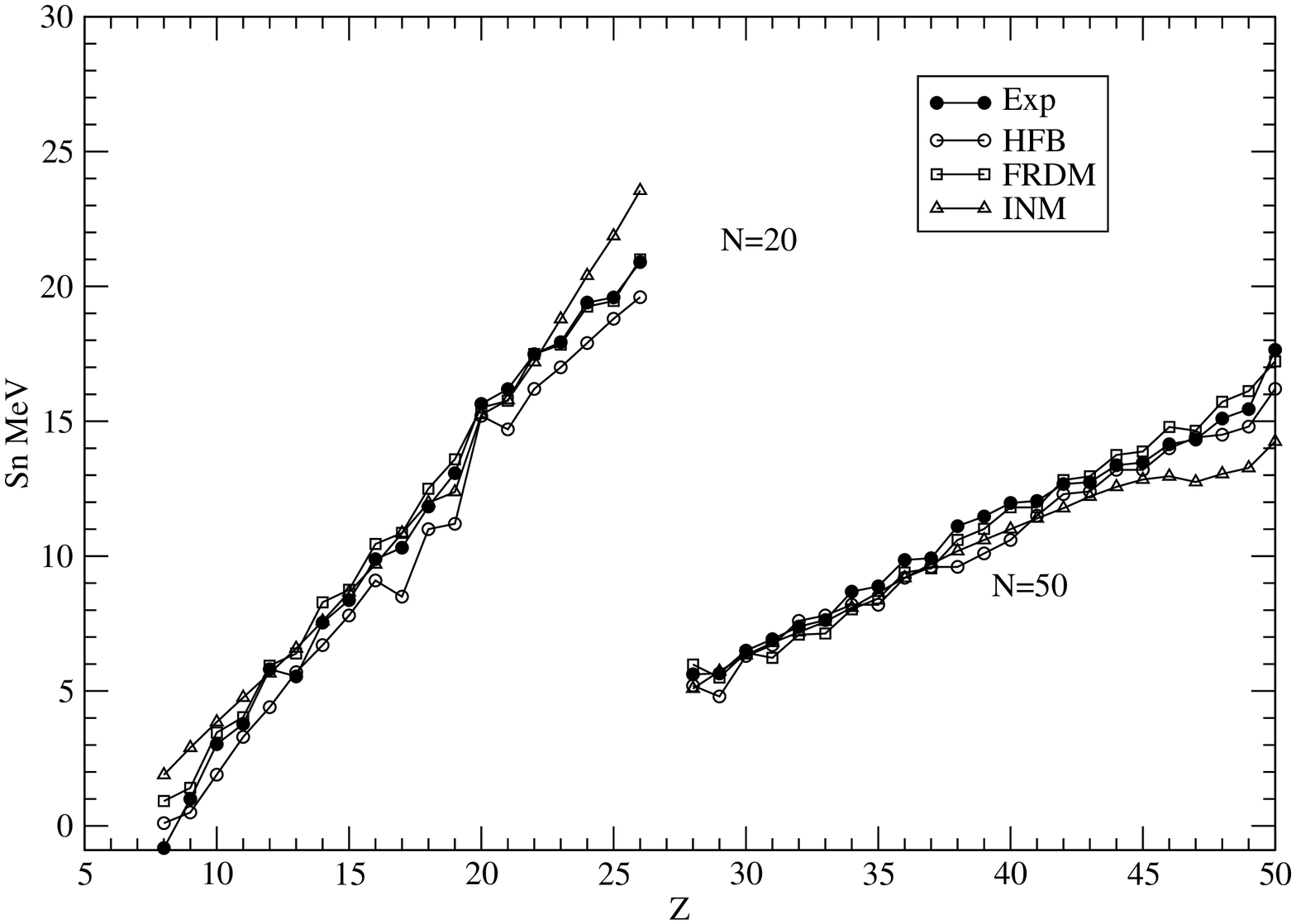}
\caption{ $ S_{N} $ vs Z for fixed N=20 and 50}
\end{figure}

\newpage

\begin{figure}[ht]
\centering
\includegraphics[totalheight=0.7\textheight, angle=-90]{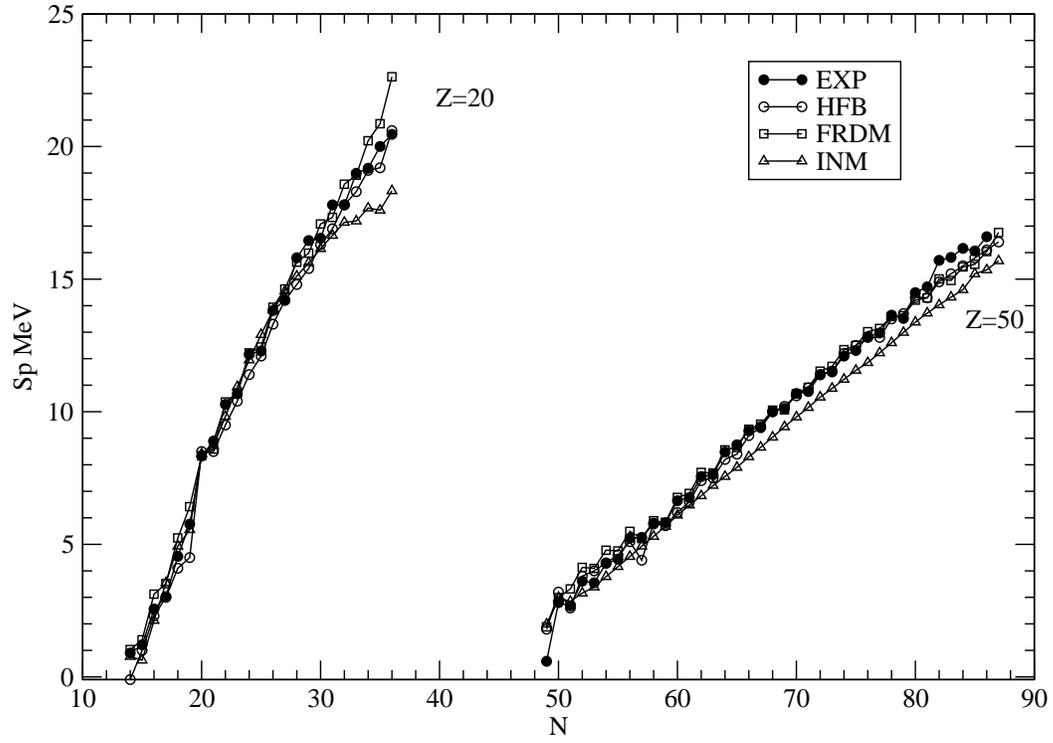}
\caption{ $ S_{p} $ vs N for fixed Z=20 and 50}
\end{figure}

\newpage

{\bf References}

\vspace{.2in}

1. S. Moszkowski, {\it Encyclopaedia of Physics}, Ed. S. Flugge et. al.,
   Springer Verlag. Berlin 1957, p. 411

2. G. Audi, A. H. Wapstra and C. Thibault, {\it Nucl. Phys.},
   {\bf A 729} (2003) 337

3. P. Moeller, J. R. Nix and K.-L. Kratz,
   {\it At. Data Nucl. Data Tables}, {\bf 66} (1997) 131

4. S. Goriely, F. Tondeur and J. M. Pearson,
   {\it At. Data Nucl. Data Tables}, {\bf 77} (2001) 311

5. R. C. Nayak and L. Satpathy, {\it At. Data Nucl. Data Tables},
   {\bf 73} (1997) 213

6. F. Coester, S. Cohen, B. D. Day and C. M. Vincent
   {\it Phys. Rev.}, {\bf C 1} (1970) 769

7. G. Auletta, {\it Foundation and Interpretation of Quantum Mechanics},
   World Scientific Press, Singapore (2000)

8. P. Ghose, D. Home and G. S. Agarwal, {\it Phys. Lett.}, {\bf A 153}
   (1991), 403; {\it Phys. Lett.}, {\bf A 168} (1992) 95

9. Y. Mizobushi and Y. Ohtake, {\it Phys. Lett.}, {\bf A 168} (1992) 1

10. S. S. Afshar, E. Flores and K. F. McDonald, {\it Found. Phys.},
    {\bf 37} (2007) 295; also S. S. Afshar - quant-ph/030503

11. J. Clarke et. al., {\it Science}, {\bf 239} (1988) 992

12. V. G. Palmieri, R. Ruggiero and P. Silvestrini, {\it Physica},
   {\bf B 284-288} (2000) 593

13. D. G. McDonald, {\it Science}, {\bf 247} (1990) 177

14. J. S. Bell, {\it Speakable and Unspeakable in Quantum Mechanics},
    2nd Ed., Cambridge Univ. Press, Cambridge, UK (2004)

\end{document}